\documentclass[a4paper]{article}

\usepackage{INTERSPEECH_v2}

\title{Putting a Face to the Voice: Fusing Audio and Visual Signals Across a Video to Determine Speakers}
\name{Ken Hoover, Sourish Chaudhuri, Caroline Pantofaru, Malcolm Slaney, Ian Sturdy}
\address{Google Inc.}
\email{\{khoover, sourc, cpantofaru, malcolmslaney, sturdy\}@google.com}

\begin{document}
\ninept
\maketitle
\begin{abstract}
In this paper, we present a system that associates faces with voices in a video by fusing information from the audio and visual signals. The thesis underlying our work is that an extremely simple approach to generating (weak) speech clusters can be combined with visual signals to effectively associate faces and voices by aggregating statistics across a video. This approach does not need any training data specific to this task and leverages the natural coherence of information in the audio and visual streams. It is particularly applicable to tracking speakers in videos on the web where {\it a priori} information about the environment (e.g., number of speakers, spatial signals for beamforming) is not available. We performed experiments on a real-world dataset using this analysis framework to determine the speaker in a video. Given a ground truth labeling determined by human rater consensus, our approach had ~71\% accuracy.
\end{abstract}

\section{Introduction}
\label{sec:intro}

In this work, we present an approach that automatically connects audio and faces in unstructured videos. We want to identify the face that corresponds to each speech signal, or conversely the speech signal that is produced by each face in a video. This problem is challenging due to the wide diversity of possible speakers, the possibility of multiple people being present, speakers being off screen, and faces turned at inconvenient angles to the camera. Our approach works without prior information or constraints on the number of people present or the structure of the conversations.

This task is particularly important as we seek to understand the overwhelming numbers of videos now available online. The growth in web-video content has led to the research toward techniques that can understand video content at scale \cite{Ton14, Abu16, Her17}, spanning efforts to identify, track and characterize semantic dimensions of content in videos (e.g. object and person detection/tracking, action recognition, event detection) \cite{Sar09,Geb15,Sch15}.

We describe an approach that makes use of today's high-quality facial-identification systems, along with speaker-diarization techniques, in order to match voices and faces. We demonstrate that we can take advantage of the multiple modalities to effectively combine robust visual signals with relatively weak speech diarization models to build a recognition system that identifies speakers in a video, including realizing they might be off-screen. Note, in this work we use the words audio, speech, voice, visual and face to represent unimodal information, while the word speaker denotes one audio and facial combination. 

One possible solution is to note the connection between movements of a face and the speech signal. FaceSync \cite{Sla00} did this by using CCA to model the correlation between pixels of the lower face and the audio signal. One could measure this correlation across all faces and then pick the face that has the highest correlation with the voice. But this system requires careful alignment, in three-dimensions, of the face to a reference image. Our approach uses a simpler but more robust signal: the presence of the same face across many speech segments to make a connection.

Advances in the computer vision community around neural network-based face detection and recognition technology has resulted in the development of technology that is robust and ubiquitous. Thus we can use face recognition\footnotemark to tie together the audio from multiple points in a video.
Here face recognition refers to the ability to determine which faces correspond to the same person (face clustering), and not the ability to identify a face as one of a set of reference images. Our approach does not require a reference library of face images.

A significant benefit of our approach is that it requires fewer assumptions about the data: we do not need to know the number of speakers \cite{Shu11}; nor do we need to assume the lower half of the video corresponds to the mouth \cite{Mon11}; nor that the face is onscreen and audio tracks are synced \cite{Hu15}; nor that we have access to the script \cite{Eve06}. 

The contributions of this work can be summarized as follows. First, we demonstrate an effective approach to combining weak audio and visual\footnotemark signals to connect faces with voices while using no labeled data for this task. Second, improving upon existing work, our approach is designed to tackle perceptual corner cases such as understanding when a speaker is offscreen. Third, we place no restrictions on the recording environment (i.e our approach works on top of any video with a mono audio stream, and does not use spatial information), the conversational structure and participants entering or leaving the recording. Finally, our approach is practical for videos on the open web as it does not require prior knowledge (training samples of faces or voices) of the individuals that are present. 
\footnotetext{The visual signal is ``weak" in the sense of identifying the active speaker, yet it is state of the art at detecting faces.}

Section \ref{sec:approach} describes our approach and system in detail and Section \ref{sec:experiments} describes our experimental framework, the human rating task to evaluate performance, and presents a discussion of our results. We conclude with thoughts on future work in Section \ref{sec:conclusions}.

\section{Proposed Approach}
\label{sec:approach}

In this paper, we demonstrate that a combination of weak audio and visual signals aggregated across a video serves as a strong signal for the task of connecting a voice to the face of a speaker. 
Our approach relies on the intuition that video content is shot with the camera attempting to focus on the speaker. During a speaker's turn, the camera is likely to show their face for some time, and the likelihood that this happens at least once over the course of multiple turns is extremely high. This intuition, combined with (the near-perfect) state-of-the-art face clustering enables us to effectively connect faces with voices, even though we have not {\it seen} or {\it heard} the specific speaker before. 

\begin{figure}[t]
  \includegraphics[width=\linewidth]{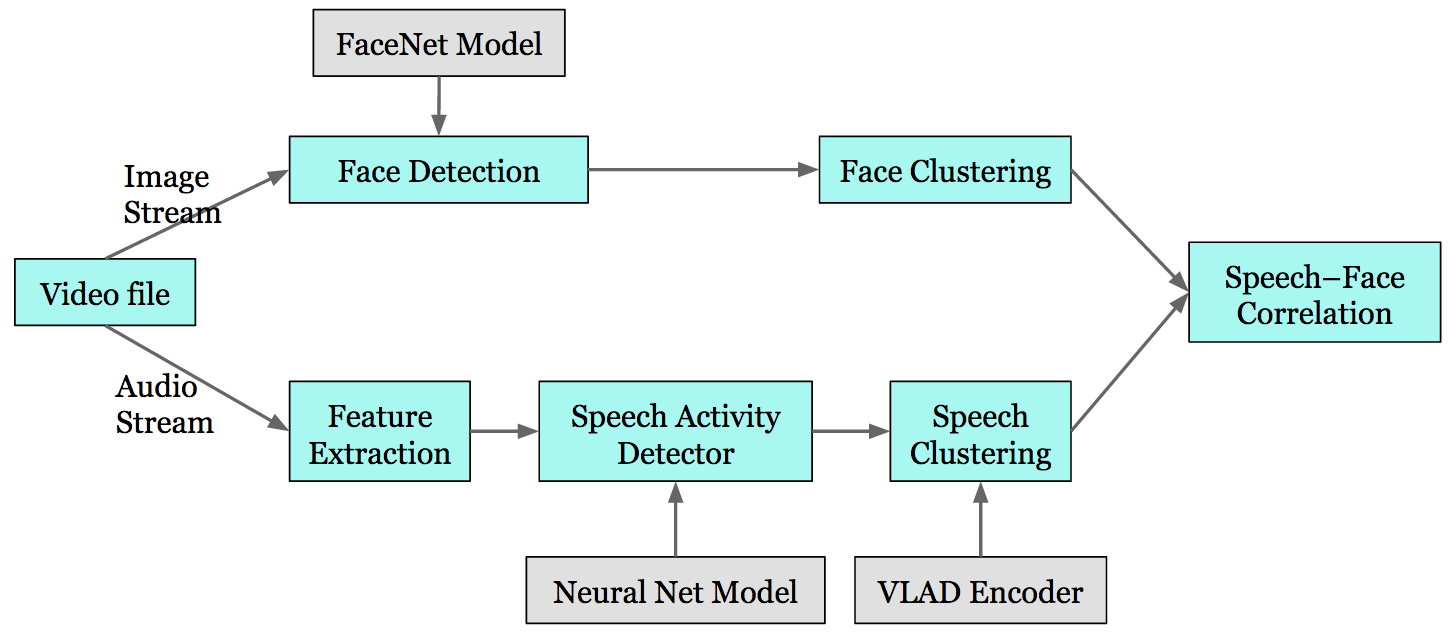}
  \caption{Overview of the processing pipeline at inference time. Grey components represent pre-trained models for the pipeline.}
  \label{fig:pipeline}
\end{figure}

A block diagram of our pipeline is presented in Figure \ref{fig:pipeline}. The upper branch shows the image stream processing, the lower branch the audio stream processing and the signals are combined in the right-most block. We note that the key novelties of our approach are in the {\it Speech Clustering} and the {\it Speech-Face Correlation} blocks which we describe in the next subsections. 

\subsection{Image Stream Processing}
\label{subsec:imageprocessing}
The only signal from the image stream used in this work involves detecting and clustering faces. Our system uses a pre-trained FaceNet system \cite{Sch15} that detects faces and maps them to a 128-dimensional embedding. This convolutional neural network model was trained using triplets of matching and non-matching examples and the final embedding represents closeness in Euclidean space based on the similarity between the facial images. FaceNet reports a recognition accuracy of 99.63\% on the Labeled Faces in the Wild dataset and 95.12\% on the YouTube Faces DB. In our system, we track and cluster faces using the FaceNet embeddings. Now, given the 128-dimensional FaceNet embedding representation for {\it all} faces found in the video, we call all the face points within a neighborhood determined by a fixed threshold a face cluster. This face cluster captures the visual appearance of a single person across the entire video.

\subsection{Audio Stream Processing}
\label{subsec:audioprocessing}

\begin{figure}[t]
  \includegraphics[width=\linewidth, height=12cm]{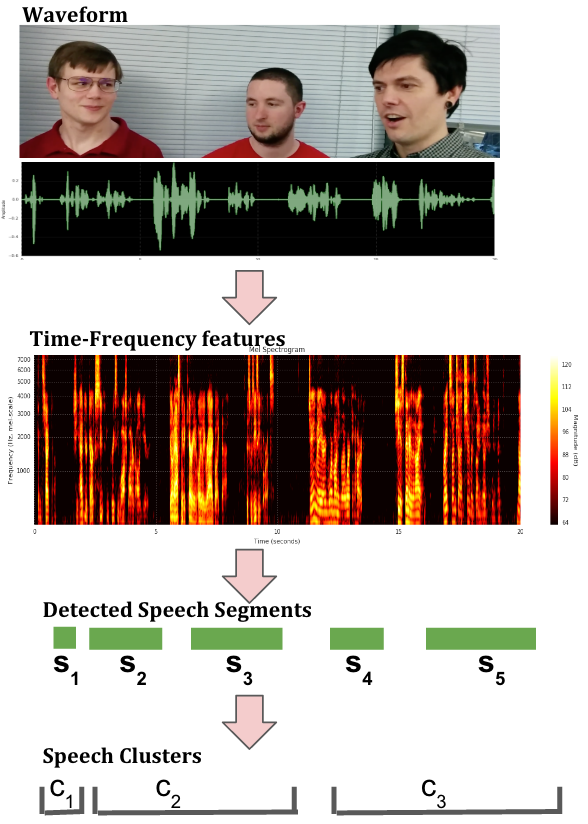}
  \caption{An illustration of the stages in the audio pipeline.}
  \label{fig:audiofig}
\end{figure}

The audio analysis pipeline is illustrated in Fig \ref{fig:audiofig}. The audio is converted to the frequency-domain for input to the downstream stages. The speech detection stage (Section \ref{subsubsec:sad}) generates speech segments, indicated by the green boxes in Fig \ref{fig:audiofig}. We assume that the resulting speech segments are uttered by a single speaker, eliminating the need for a speaker turn-change decision. We then (under-)cluster these segments (Section \ref{subsubsec:vlad}), giving us speech clusters which likely belong to a single speaker. While better turn-change detection will improve performance \cite{Shu13}, it is orthogonal to the issue being investigated here: that combining simple audio and visual information creates a strong system for determining the speaker. The final analysis step of Speech-Face correlation enables us to balance the relatively weak (i.e. conservative) speaker-diarization system with strong visual cues from the detected faces.

\subsubsection{Speech-Activity Detector}
\label{subsubsec:sad}
We use a time-frequency representation of the audio and a neural network-based detector to detect speech activity\footnotemark.
\footnotetext{The model was trained to be sensitive to conversational speech, and distinguishes it from other vocal activity such as singing and laughing.} 
The model predicts whether speech is present at the rate of 100 predictions per second. We then smooth these dense predictions into segments by requiring that a speech segment be at least 1 second long, and that the gap\footnotemark ~ between consecutive segments are at least 0.25 seconds long, and segments with smaller gaps are collapsed into one.\footnotetext{While the speech literature often distinguishes between gaps and pauses (the former has different speakers on either end, the latter has the same speaker), we do not make such a distinction.}

\subsubsection{Speech (Speaker) Clustering with VLAD}
\label{subsubsec:vlad}
The task of clustering speaker turns has received considerable attention in the context of speaker diarization, which tries to answer the question: ``who spoke when" \cite{Tra06, Moa12, Ang12}. Our work differs from the prior diarization research in three ways. First, our intended domain of application is web videos where carefully labeled datasets do not exist and would be expensive to obtain. Second, videos are recorded in a variety of conditions and control or instrumentation of the space that would enable localization signals (i.e. \cite{Geb15}) is infeasible. Finally, the content is free form with participants entering and exiting the scenes, with no further information available {\it a priori}. 

The combination of the issues above presents significant challenges to diarization, and especially the clustering phase, which typically involves hierarchical agglomerative clustering and relies on heuristics to determine when to stop clustering. We believe that visual information presents the best opportunity for determining the correct clustering, which is the approach this work takes. The audio clustering step, therefore, chooses a conservative stopping criterion that results in under-clustering, {\it i.e.} more clusters than speakers. 

We start by assuming that each speech segment from the speech-activity detector belongs to a singleton cluster, and compute a (fixed size) embedding for this (variable length) segment.  We use a Vector of Locally Aggregated Descriptors (VLAD) technique \cite{Jeg10}, from the image-processing world, to represent each segment (or combinations of segments) as a point in the embedding space. VLAD, like i-vectors, is based on local vector displacements from cluster centers that are computed from a large speech database. We then use a hierarchical agglomerative clustering \cite{Rok05} algorithm with complete linkage \cite{Def77} on the speech segments, and the stopping is determined by a similarity threshold, that is set high to ensure conservative merging of clusters. In Figure \ref{fig:audiofig}, the clustering stage generates 3 clusters where perhaps there should have been 2 because the first 3 speech segments have the same face in the video. Nonetheless, the under-clustering is corrected in the next stage, where the correlation of these clusters with faces accurately identifies the speaker. 
We refer to the clusters generated in this manner as {\it speech clusters}, henceforth.

\subsection{Mapping Voices to Faces}
\label{subsec:mappingfacevoice}

\begin{figure}[t]
  \includegraphics[width=\linewidth]{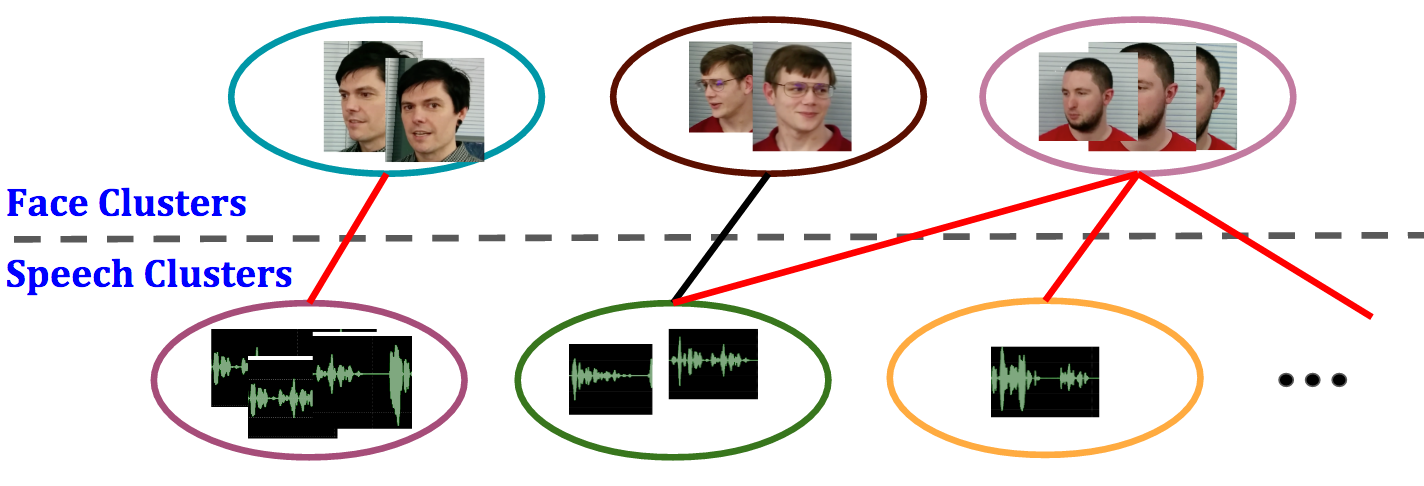}
  \caption{Connecting speech segments to faces. Edges exist when a face and speech cluster co-occur. Black edges represent co-occurrence, red represents the chosen connection. Only one face is seen during the first speech cluster, so it is easy to assign that face to that voice. The second speech cluster co-occurs with two different faces. The third face cluster above is assigned to this second speech cluster because its faces co-occur more often on-screen during the time of this speech cluster.}
  \label{fig:mapping}
\end{figure}

The final phase combines the information extracted from the audio and visual streams to determine the speaker for each speech segment. Recall that we expect face clusters to be perfect (each face cluster contains images of only one individual and every instance of that individual) and that speech clusters contain exactly one speaker (because of conservative agglomeration). We determine, for each speaker cluster, which face cluster co-occurs with it most often. Thus for any given speech cluster we choose the face cluster that appears in the greatest number of video frames during the temporal extent of the speech/speaker cluster. Since multiple speech clusters can map to the same face cluster, this effectively functions as the final stage of speech clustering, and in the process assigns a face to the speech. Figure \ref{fig:mapping} shows an example of the matching process. 

\section{Experimental Results}
\label{sec:experiments}

We experimented with different front-end features and data sources for training the VLAD model (Section \ref{subsec:vladexpt}) and discuss results from running the full pipeline (Section \ref{subsec:mapdiscuss}).
For our experiments, we used FaceNet \cite{Sch15} with a threshold of 0.85 on normalized embedding similarities.

\subsection{VLAD Model Experiments}
\label{subsec:vladexpt}

\begin{figure}[t]
  \includegraphics[width=0.9\linewidth]{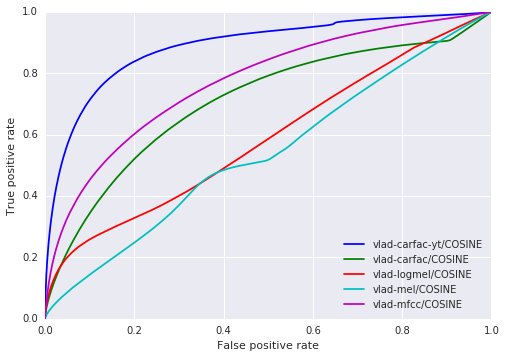}
  \caption{ROC curves for the VLAD system with four different features trained on two different sets of speech data (LibriVox vs. YouTube) on the task of distinguishing same v/s different speakers (cosine similarity was the similarity metric used).}
  \label{fig:rocplot}
\end{figure}

For the audio pipeline, we experimented with four different front-end feature representations: a Mel filterbank \cite{Ste37} was used to generate the Mel spectra, log-Mel and Mel-Frequency Cepstral Coefficients and a CARFAC filterbank \cite{Lyo11} front-end. To select the best one, we trained the VLAD encoder model with each feature, and evaluated its performance on a dataset of speaker-labeled utterances (LibriVox \cite{Lvo}) to determine how well the embeddings were able to determine same v/s different speakers. The evaluation set consisted of 100,000 pairs of LibriVox audio clips (average length 3 seconds, from 1200 unique speakers) with the label {\it same} or {\it different}. For each feature type, we generated ROC curves (see Figure \ref{fig:rocplot}) by changing the threshold for the boundary between {\it same} and {\it different}. We contrasted performance when computing 128 VLAD clusters on LibriVox data versus speech segments from YouTube. Figure \ref{fig:rocplot} displays a set of the ROC curves and shows the VLAD model trained on YouTube data with CARFAC features gave the best speaker discrimination.
We used the CARFAC feature in our audio pipeline.
We trained the VLAD model on $\sim$2M speech segments from $\sim$10K YouTube videos. No labels, keywords, descriptions or speech transcripts were used from the metadata associated with any of these videos.

\subsection{Results of Mapping Voices to Faces}
\label{subsec:mapdiscuss}

\begin{figure*}[t]
\begin{minipage}{0.24\linewidth}
  \centering
  \centerline{\includegraphics[width=\columnwidth, height=3cm]{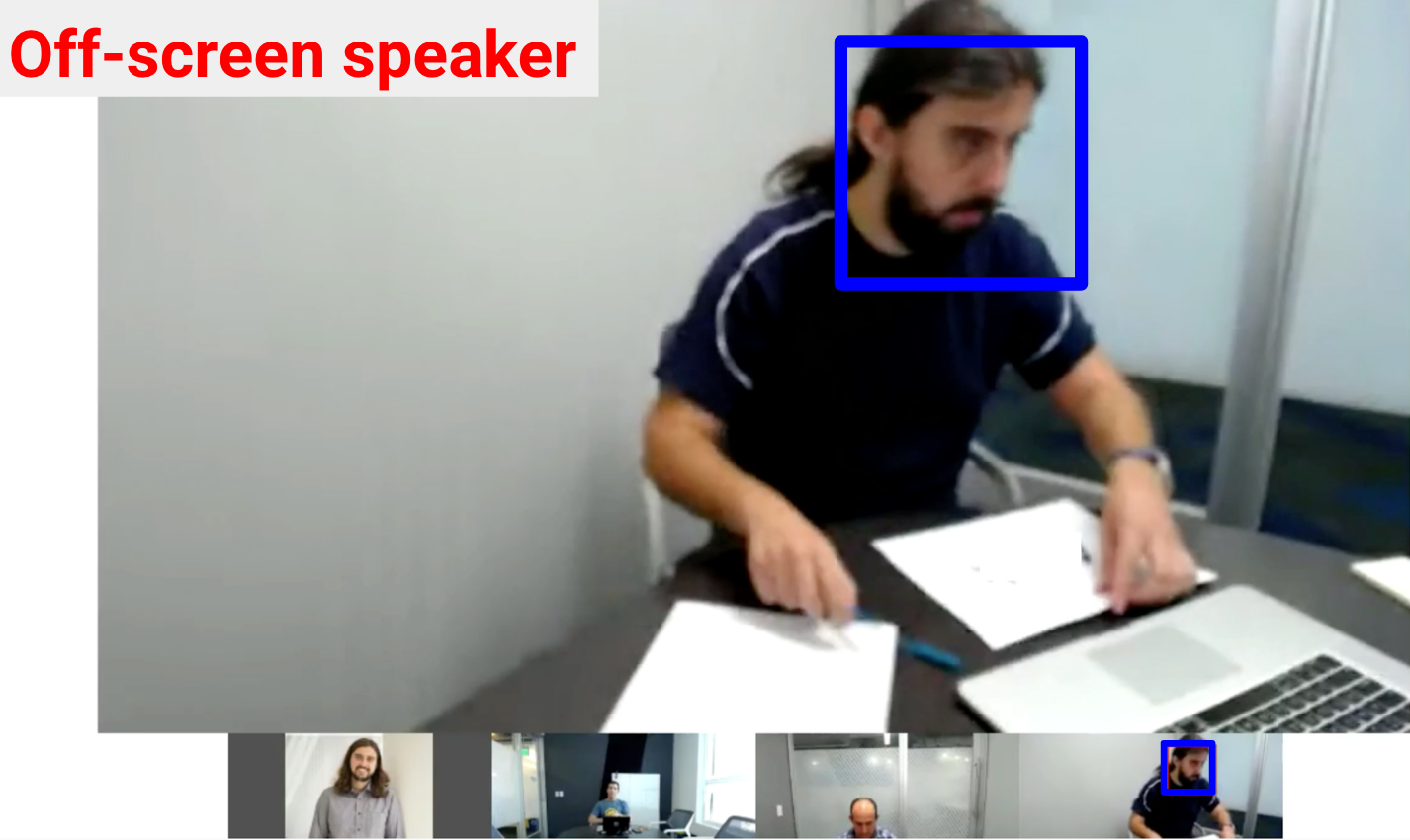}}
\end{minipage}
\hfill
\begin{minipage}{0.24\linewidth}
  \centering
  \centerline{\includegraphics[width=\columnwidth, height=3cm]{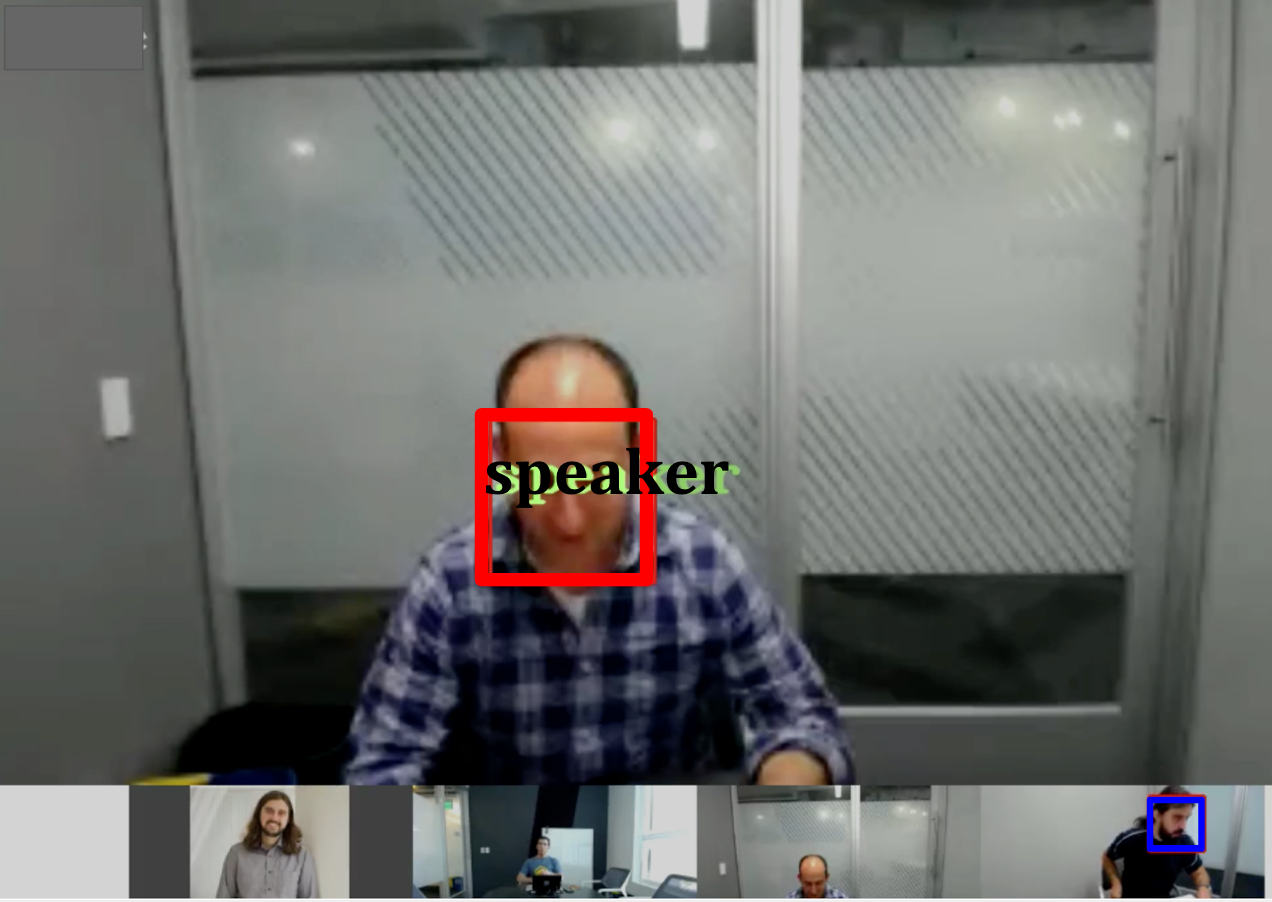}}
\end{minipage}
\hfill
\begin{minipage}{0.24\linewidth}
  \centering
  \centerline{\includegraphics[width=\columnwidth, height=3cm]{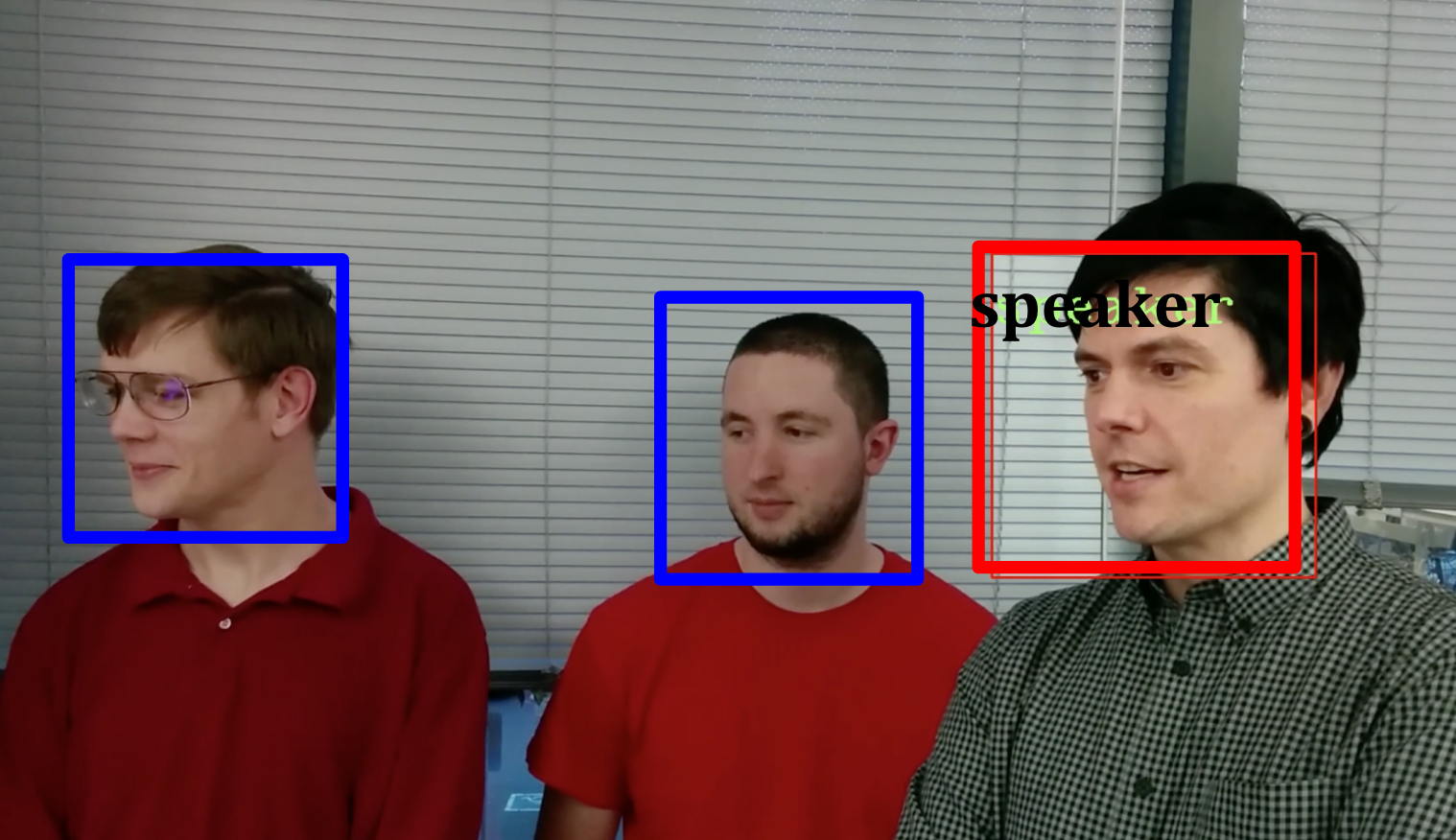}}
\end{minipage}
\begin{minipage}{0.24\linewidth}
  \centering
  \centerline{\includegraphics[width=\columnwidth, height=3cm]{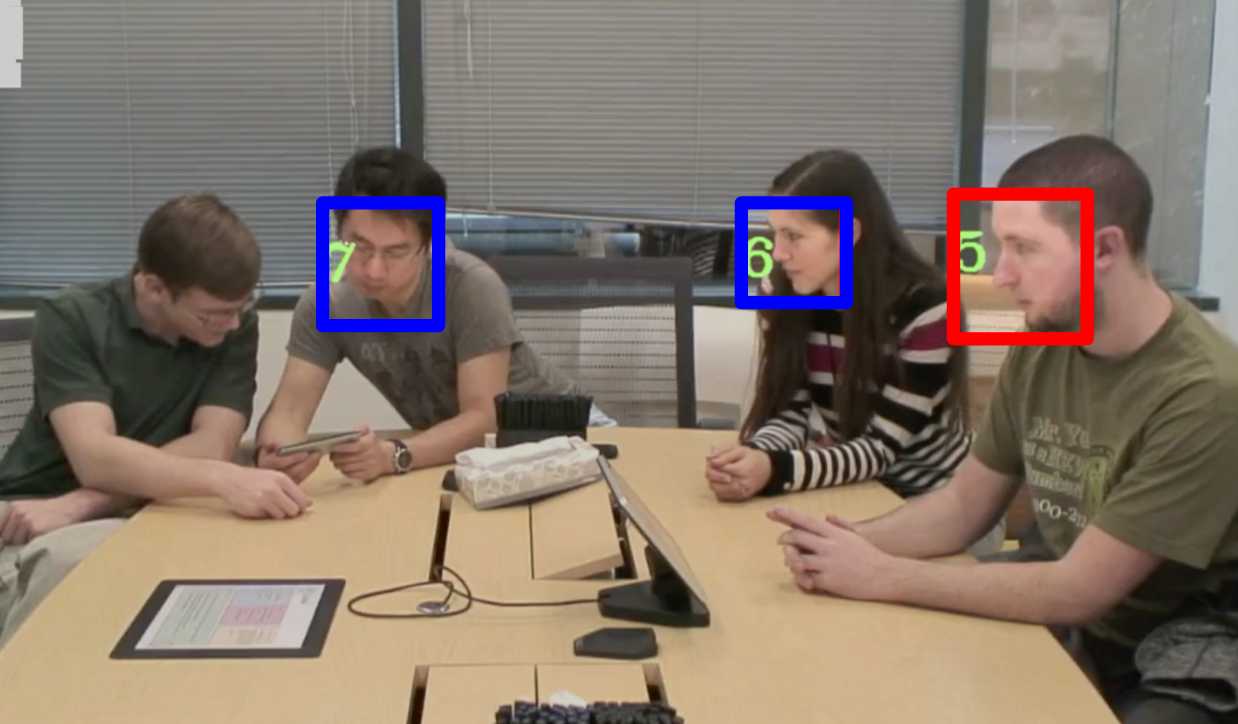}}
\end{minipage}
\caption{Exemplar results from our pipeline, with the first 3 being instances of correctly determined speakers and the last an instance of an error. A blue box around a face indicates a detected face only, while a red box indicates that that detected face was determined to be the speaker. From left to right: {\bf (a)} \& {\bf (b)} Shots from within a few frames of each other, where the system correctly determines the speaker is offscreen (leftmost image) before the shot shifts to the speaker who is correctly identified (image second from left). {\bf (c)} The speaker is correctly tagged in the presence of 2 other faces. {\bf (d)} Actual speaker is on the far left, and the combination of a small number of pixels and tilted face pose resulted in the speaker's face not being detected.}
\label{fig:resgood}
\end{figure*}

Given the pipeline from Figure \ref{fig:pipeline} and an input video, we can determine the speaker at any point in the video. Figure \ref{fig:resgood} shows some example results from our pipeline, including correctly determining offscreen speakers, identifying the correct face as speaker in shots with multiple faces, and an error case. 

Since ground-truth labeled data for videos in the wild aren't available, we quantify performance using a human rating task. This task was set up as a verification task, where a rater was presented with a video clip 10--15 seconds long and the video had the label {\it SPEAKER} on the bounding box over the face determined to be the speaker (as shown on the examples in Fig \ref{fig:resgood}). If the system determined there was speech but the speaker was not visible on-screen, it shows the tag {\it OFF-SCREEN SPEAKER} as on the left panel in Figure \ref{fig:resgood}.


We evaluated performance on a set of 400 randomly sampled YouTube videos. We presented 3558 clips totaling $\sim$12 hours from these videos to a pool of human raters. For each clip, we asked three raters if the presented speaker labeling was correct: that is, when the {\it Speaker} or {\it Off-Screen Speaker} tag appeared, it was accurate, and that such a tag correctly labeled all speech activity. The rater could mark the clip as Correct, Incorrect, Partially Correct or Unsure. The rater used the label Partially Correct when part of the segment had the correct speaker identified while another part had an incorrect speaker identified, and Unsure when the rater couldn't tell, e.g. multiple people with backs to the camera, possible voice-over. For 6 out of the 3558 segments, a single rater selected the Unsure option.

To understand the rating difficulty of the task, we spot-checked the rated results for correctness, and computed Fleiss' kappa on the data to measure of inter-rater agreement \cite{Fle71}. The kappa on this speaker label verification task was 0.732 (where 1 is perfect agreement, and 0 or less is no agreement above chance), indicating significant agreement above chance. 

\begin{figure}[t]
  \includegraphics[width=\linewidth, height=5.3cm]{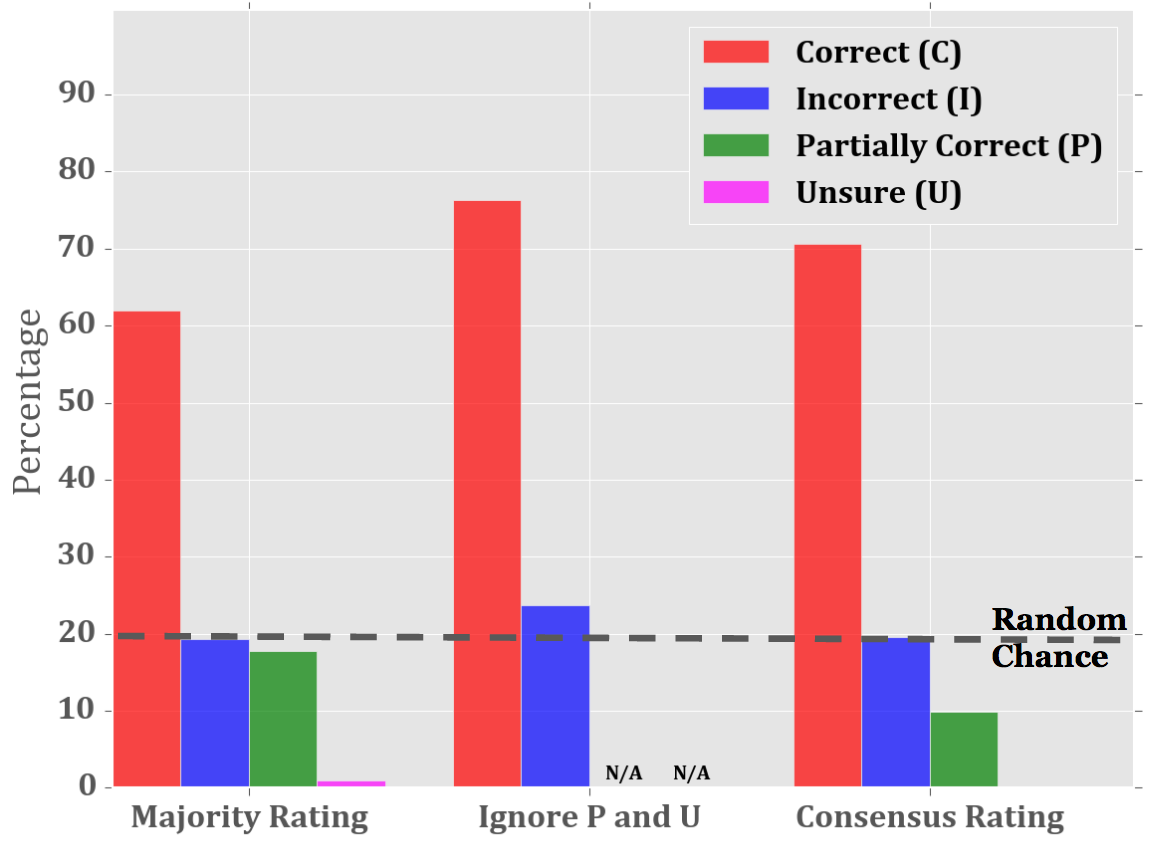}
  \caption{Speaker association results with different aggregations of ratings from 3 raters, compared with random chance (indicated by the dotted line) performance.}
  \label{fig:results}
\end{figure}

The final results from the human rated segments depend on how the 3 ratings per clip are aggregated and how the Partially Correct and Unsure ratings are counted. Results by percentage are provided corresponding to different rating handling and aggregation schemes are in Figure \ref{fig:results}. Note that the annotated video segments had an average of 13 faces per video, although on average only 5 speakers accounted for over 85\% of screentime, and we use the lower number 5 to set the baseline probability of randomly selecting the right face as 20\%. 

Figure \ref{fig:results} shows that performance of this approach is considerably above chance performance on this task, but also indicates there is room for improvement. To understand the errors better, we scanned the data for patterns in the errors where a particular error mode stood out. In 65\% of the error cases, there was a violation of the assumption that a speech segment was uttered by a single speaker, {\it i.e.} there were a number of cases of multiple speakers speaking one after the other without a break in speech activity resulting in a single speech segment. Since improving speaker turn change detection wasn't a focus of this work, we anticipate further gains in the future with a state-of-the-art speaker turn-change detector. 


\section{Conclusions and Future Work}
\label{sec:conclusions}

This paper presents an approach for joint audiovisual analysis of videos to determine the voice and face of a speaker at any point in the video. It needs a pre-trained face and speech detector (and high-accuracy versions of both are now easily available), but doesn't need any other training data. This is the first system, to the best of our knowledge, to develop an explicit multimodal notion of speakers that can understand sufficient context to identify the speaker as offscreen when they aren't visible, and significantly outperforms random chance on this task. We expect that improving the signals in each of the modalities will lead to further gains in performance: specifically, using motion information in the visual domain to predict a visual-speaking signal, and improving the audio pipeline with a speaker turn-change detector. 

While the determination of the speaker is useful for many applications, we believe this perceptual framework is powerful beyond this task. Audiovisual correlations will help understand scene structure, enable better understanding of human interactions, adapt speech recognition models to active speakers. Our labeling pipeline in combination with automated methods should enable generation of large-scale labeled datasets from videos in the wild and will improve state-of-the-art for audio-only and visual-only applications.

\bibliographystyle{IEEEtran}
\bibliography{is2017bib}

%
%
%
%
%

\end{document}